\documentclass[iop,apj]{emulateapj}
\usepackage{bm}
\usepackage{color}
\usepackage[caption=false,position=top,labelformat=empty]{subfig}
\usepackage{graphicx}
\usepackage{amsmath}
\usepackage{tabularx} 
\usepackage{natbib}
\newcommand{\mpo}[1]{{\color{black}#1}}

\begin{document}
\title{Time variability of TeV cosmic ray sky map}
\author{Rahul Kumar$^{1}$}
\altaffiltext{1}{Department of Astrophysical Sciences, Princeton University, Princeton, NJ 08544, USA}
\author{Noemie Globus$^{2}$}
\altaffiltext{2}{KIPAC, PO Box 20450 MS29, Stanford, Palo Alto, CA 94309, USA}
\author{David Eichler$^{3}$}
\altaffiltext{3}{Department of Physics, Ben-Gurion University, Be'er-Sheba 84105, Israel}
\author{Martin Pohl$^{4,5}$}
\altaffiltext{4}{DESY, 15738 Zeuthen, Germany}
\altaffiltext{5}{Institute of Physics and Astronomy, University of Potsdam, 14476 Potsdam, Germany}

\begin{abstract}
The variation in the intensity of cosmic rays at small angular scales is attributed to the interstellar turbulence in the vicinity of the Solar system. We show that {a turbulent origin of the small-scale structures implies that} the morphology of the observed cosmic-ray intensity skymap varies with our location in the interstellar turbulence. The gyroradius of cosmic rays is shown to be the length scale associated with {an observable change in the skymap over a radian angular scale. The extent to which the intensity \mpo{at a certain} angular scale varies is proportional to the change in our location with a maximum change of about the amplitude of intensity variation at that scale in the existing skymap.} We suggest that for TeV cosmic rays a measurable variation could occur over a time scale of a decade due to the Earth's motion through the interstellar medium, if interstellar turbulence persists down to the gyroradius, \mpo{about $00\ \mu\mathrm{pc}$ for TeV-ish cosmic rays}. Observational evidence of the variability, or an absence of it, could provide a useful insight into the physical origin of the small-scale anisotropy.
\end{abstract}
\maketitle

\section{Introduction}
A meticulous measurement of energy and arrival direction of individual primary cosmic ray (CR) nuclei has revealed that the intensity of CRs in the TeV band is remarkably isotropic. The measured degree of deviation from isotropy, about few parts per ten thousand, appears to be much smaller than the implied level of anisotropy by the standard diffusive propagation models of Galactic CRs, which are otherwise effective in reproducing several observables, such as the spectrum and composition \citep[see e.g.][]{Strong2007,Blasi,Pohl2013}. However, the consideration of anisotropy in the spatial diffusion and an inhomogeneous distribution of cosmic-ray sources in the spiral arms can potentially reduce the tension between the observation and the propagation models \citep{David2013,Rahul2014TeV,Rahul2014EeV,EvoliPRL,David2016}. 

In recent years, a large number of TeV-band CR events registered by modern observatories have manifested small but significant variation in the flux at all resolved angular scales \citep{Amenomori439,SuperPRD,MilagroPRL,ARGO,HAWC2014}. Although the dipole anisotropy in the measured intensity is understood to be due to asymmetry in the distribution of CR sources, the origin of anisotropy in the intensity at the small angular scales remains unclear. It has been suggested that these patchy structures in the CR intensity skymap may be a combined effect of the prevalent interstellar magnetic turbulence in the vicinity of the Solar system and a spatial inhomogeneity in CR density \citep{Gwenael2012,Ahlers2014,Ahlers2015}. Alternatively, it has been suggested that they could possibly be due to very local alteration of the CR phase-space distribution by heliospheric structures \citep{IbexAnis,Ming2014,Drury2013,Paolo2013,Paolo2016}. 


In this paper, we suggest that if the \mpo{CR anisotropy} at small angular scales is mainly due to the local turbulence, then the structures in the CR \mpo{arrival distribution} of a given energy band should be variable on a time-scale determined by gyroradius-scale shift in the position of the Earth with respect to the turbulence. At a velocity of about 20 km s$^{-1}$, the solar system traverses $\sim40$ AU ($\sim 2\, \cdot 10^{-4}$ pc) in approximately 10 years\citep{Witte2004,ibex2009}, which is comparable to the gyroradius of TeV CRs close to our solar system, \mpo{$r_g(E) \sim 3\, \cdot10^{-4}$ pc for protons and 3 $\mu$G as the estimated strength} of the magnetic field in the local interstellar medium \citep{ibexB}. For TeV CRs the variation due to a change in the location of the Solar system is expected to cause a decadal-scale variation. 

The HAWC observatory operating in the Northern Hemisphere reports a significant change in the intensity skymap of few-TeV CRs\citep{HAWC2014} in relation to the data accumulated by its predecessor, the Milagro experiment \citep{MilagroPRL}. Specifically, it reports a change in the intensity and location of one of the \mpo{intensity excesses and the} appearance of an entirely new excess region at TeV energies. While the discrepancy is suggested to be due to a difference in the median energy of the \mpo{CR events recorded by} these two observatories, it is likely that part of the discrepancies are of physical origin due to change in the location of the Earth over the operational period of these observatories. In the following we present the numerical method used for estimating CRs flux in differential angular bins at the Earth. Thereafter, we compute skymaps of CRs for slightly different locations of the Solar system to illustrate the variability of \mpo{the CR arrival distribution}. 

\section{Numerical method}

We numerically integrate the equation of motion of charged particles in a model turbulent magnetic field to exemplify the chaotic nature of the cosmic-ray trajectories and the origin of small-scale anisotropy.  Formally, trajectories of CRs in phase-space are described by a doublet of vectors, $\vec{x}(t)$ and $\vec{p}(t)$, denoting the position and the momentum of a CR at time $t$. The time evolution of the trajectories is governed by $d \hat{p} / dt = \omega_c \hat{p}\times \hat{B}$; and $d\vec{x}/dt =\vec{p}/m$, where $\hat{p}$ and $\hat{B}$ are unit vectors along the momentum and the local magnetic field, respectively, $\omega_c$ is the energy dependent gyro-frequency, and $m$ is the relativistic mass. The three-dimensional \mpo{magnetic turbulence} used in our computation is represented by superposition of several linearly polarized shear Alfv\'en waves of random polarizations and phases \citep{Jokipii1999}. The wavenumbers of Alfv\'en modes are logarithmically spaced with $d \log k=0.01$ between minimum wavenumber $k_{min}=0.5/ l_c $ and maximum wavenumber $k_{max}=5\cdot 10^3/ l_c$, where $l_c$ is the correlation length. The turbulent magnetic field is assumed to follow a \mpo{spatially homogeneous} Kolmogorov scaling. The ratio of mean free path to $r_g$ for CRs of smaller $r_g/l_c$ is larger and requires larger computational time to illustrate random walk of the CR trajectories. {Therefore, majority of our analyses are performed for large rigidity CRs, i.e. $r_g/l_c \simeq 0.1$. However, in section \ref{lowEsec} we illustrate that our results are valid for low rigidity CRs as well.}  

\begin{figure}
 \includegraphics[width=.99\linewidth]{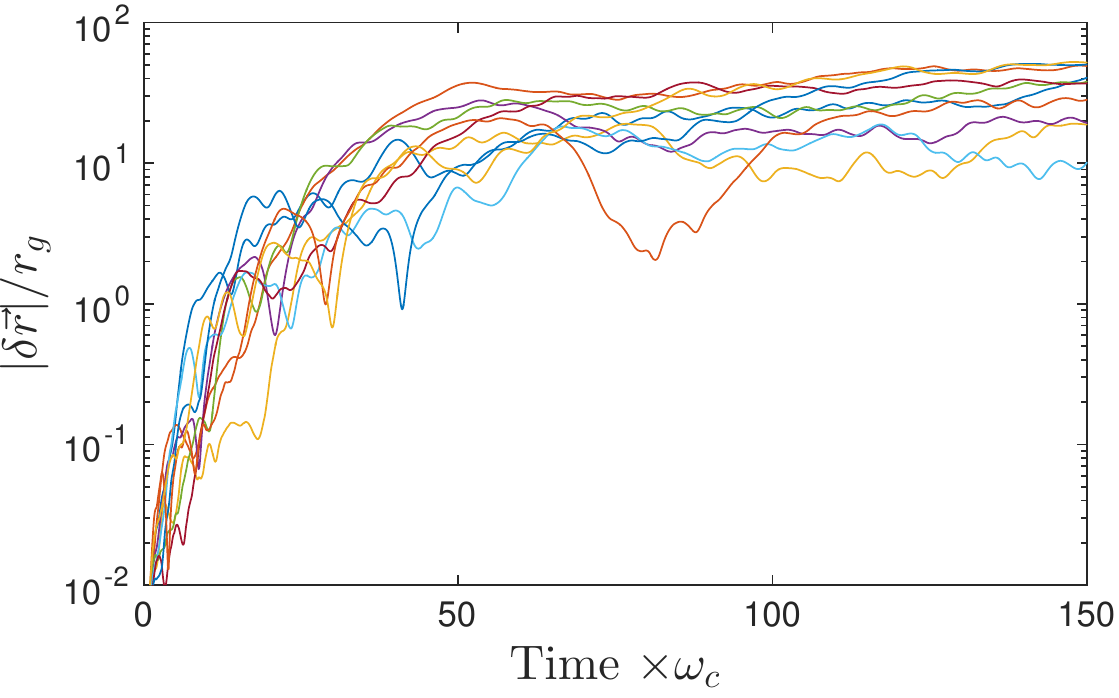}
\caption{The magnitude of spatial separation, $\delta \vec{r}$, between a pair of CR trajectories are shown as a function of time. Initially $\delta \vec{r}=0$ and $d\vec{p}=10^{-2}\vec{p}$. Each curve \mpo{represents a} pair of trajectories starting from the same location but with randomly oriented momentum vector. }
\label{correlation}
\end{figure}

Trajectories of charged particles can be decomposed into gyration around a mean magnetic field, $\vec{B_0}$, and \mpo{the variation in pitch angle (the angle between $\vec{p}$ and $\vec{B_0}$) arising from} magnetic fluctuations, $\delta \vec{B}$, perpendicular to $\vec{B}_0$, if $\delta B^2/{B}^2_0$ is much smaller than unity \citep{Eichler1987}. 
However, the randomization of trajectories cannot always be quantified in terms of pitch-angle scattering, e.g., in the absence of a coherent large-scale mean magnetic field. More generally, the correlation time of nearby points in phase-space, defined by the Lyapunov exponent, can be used to quantify the chaotic nature of the trajectories. In figure \ref{correlation} we show the time evolution of the spatial separation between few randomly selected pairs of trajectories whose initial momentums are separated by $d\vec{p}=10^{-2}\vec{p}$. The separations between the pair of trajectories first increase nearly exponentially, and the initial correlation in their momentum erodes away. After the trajectories are decorrelated, they \mpo{reflect} independent random walks in phase-space and their separation increases diffusively ($\propto t^{1/2}$). Clearly, the time over which two trajectories remain correlated depends on their initial separation in phase-space. In addition, the numerical integration of the equation of motion at each time step involves an irreducible error characterized by the error tolerance adopted by the numerical scheme. 
Therefore, numerically backtracked trajectories of individual CRs cannot be used deterministically to map the phase-space density at two different times by invoking Liouville's theorem. We note that the source of error in mapping the phase-space density is not merely due to numerical discretization, but it could also have a physical origin due to small scale ($\ll r_g$) magnetic fluctuations present in the interstellar medium \mpo{that are} not resolved by the observations or the numerical simulations.

However, correlation in the backtracked locations of multiple CRs  (at $t=-T$, say) of known arrival direction at Earth (at $t=0$) can be used to estimate correlation in their spatial origin. Specifically, we consider $n$ CRs whose arrival directions at the Earth are uniformly distributed within a two-dimensional angular bin of solid angle $d \Omega$ pointing in direction $\hat{r}$ in an arbitrary coordinate system. We numerically compute \mpo{the} position of each CR at the backtracked time $t=-T$, denoted as $\vec{x}_i(-T) \,  \forall i \in \left\{1,2,...,n\right\}$. The phase-space separation among the CRs increases with the backtracking time, and \mpo{the} initial correlation in their position and momentum is lost after $t\gtrsim \tau_c$.  
At the back-tracked time $t=-T$, we view the spatial distribution of their backtracked location $\vec{x}_i(-T)$ as a probability distribution of their origin. We use the centroid of the probability distribution to estimate the differential flux of cosmic rays, $F(\hat{r})$, within the angular bin, i.e. $F(\hat{r},-T) \propto  N_{cr}(\vec{x}_c(-T))$, where $N_{cr}$ is a presumed large-scale spatial distribution of CR number density and $\vec{x}_c(-T)$ is an arithmetic mean of all $\vec{x}_i(-T)$. 
For simplicity, we consider 
that the CR density distribution, $N_{cr}$, follows a uniform spatial gradient. 
The inhomogeneity in CR density implies that a variation in $\vec{x}_c$ with the orientation $\vec{r}$ of the angular bins would give rise to variation in the CR \mpo{arrival direction and hence their intensity}. 

The centroid $\vec{x}_c$ of the probability distribution is not altered by the uncorrelated random walks of the backtracked trajectories (for $t \gtrsim \tau_c$) in a spatially homogeneous and isotropic turbulence. 
However, the ensemble of backtracked trajectories would follow any existing large-scale magnetic field whose topology would contribute to the estimation of $\vec{x}_c$. The effect of large-scale magnetic field can be subtracted if the field lines are sufficiently ordered \mpo{for it to} almost equally affect $\vec{x}_c$ for all angular bins regardless of their orientation, $\hat{r}$. Otherwise, the probabilistic approach described here can not be reliably employed to estimate \mpo{the CR intensity}, since the convergence of $\vec{x}_c$ to a limit is not guaranteed for any arbitrary magnetic-field structure.  Moreover, the correlation time of trajectories, $\tau_c$, is larger for smaller angular bins, and for a sufficiently small angular bin decorrelation of trajectories due to growth of numerical error may be significant which would bias the convergence of $\vec{x}_c$. In any case, we impose the criteria of convergence for the choice of angular bin size which can be resolved by the numerical calculation and does not depend on the systematic error in the integration of trajectories. Specifically, we ensure that our results converge with respect to the error tolerance adopted in the numerical integration by changing the numerical tolerance and numerical integration method. The results presented here used a fourth-order accurate Bulirsch-Stoer integration scheme with relative error tolerance of $10^{-6}$ per $0.1/ \omega_c$ \citep{Press:2007:NRE:1403886}.  

\begin{figure*}
  \centering
         \subfloat[  $T= 3 / \omega_c$  ]{%
  \includegraphics[width=.25\linewidth]{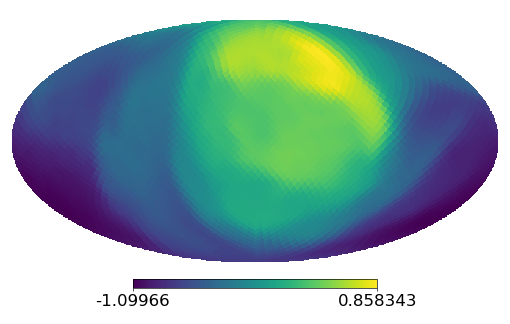}%
}\hfill
       \subfloat[ $ T= 15 / \omega_c $ ]{%
  \includegraphics[width=.25\linewidth]{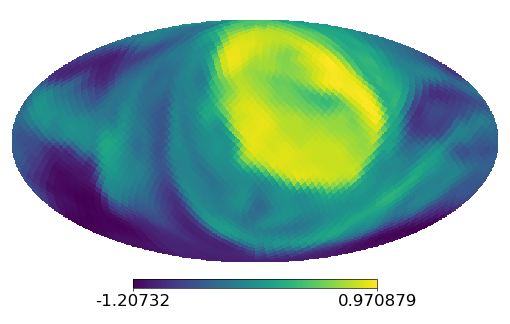}%
}\hfill
       \subfloat[ $ T=30/\omega_c $ ]{%
  \includegraphics[width=.25\linewidth]{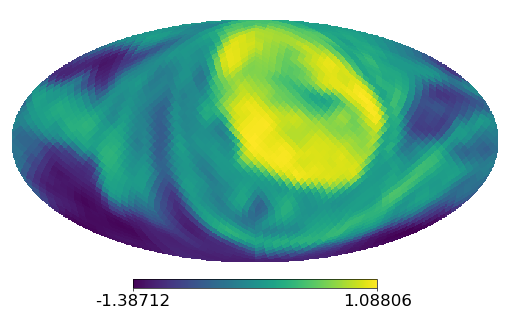}%
}\hfill
       \subfloat[ $ T=45 /\omega_c $]{%
  \includegraphics[width=.25\linewidth]{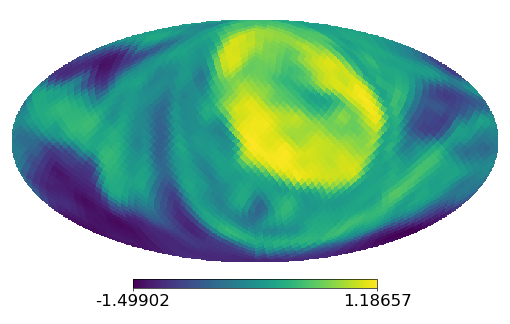}%
}\hfill

  \caption{\mpo{CR intensity skymap calculated} by backtracking cosmic-ray trajectories, here shown as a function of backtracking time. The energy distribution of CRs is a Gaussian of median $r_g/l_c=0.1$ with 10 percent \mpo{dispersion} in the energy. The method used to estimate the flux in individual angular bins is described in the text. Here and in the figure \label{skymap} a smoothing filter of 0.1 radian angular width is applied.}
\label{convergence_skymap}
\end{figure*} 

To build skymaps, we discretize the entire range of CR arrival directions into two-dimensional angular bins of equal solid angles. The shape of angular bins are determined using the Healpix \citep{healpix} program. We then estimate the CR flux in all angular bins, as described above, by backtracking a large number of CR trajectories. The skymaps of CR \mpo{intensity} constructed from the mean position of backtracked CR trajectories at four different backtracking times are shown in figure \ref{convergence_skymap}. About $10^3$ particles per angular bin of size $4\times 10^{-3}$~sr (Healpix parameter NSIDE=16) are used to estimate the CR flux in each bin. Note that the estimated CR fluxes are additive. That is to say, sum of estimated fluxes in two adjacent bins is the estimated flux in the conjoined bin. The computed intensity in all the bins are scaled such that the amplitude of the dipole anisotropy is unity. It is evident from figure \ref{convergence_skymap} that the structure of the CR \mpo{arrival distribution} as well the amplitude of small-scale anisotropy (in proportion of the dipole) converges to a limit. We consider the \mpo{converged} skymap as representative of the observed CR \mpo{intensity} for an imposed large-scale gradient in CR number density.

\section{Variation in the Skymap }

\begin{figure*}
  \centering
         \subfloat[  Reference Location  ]{%
  \includegraphics[width=.25\linewidth]{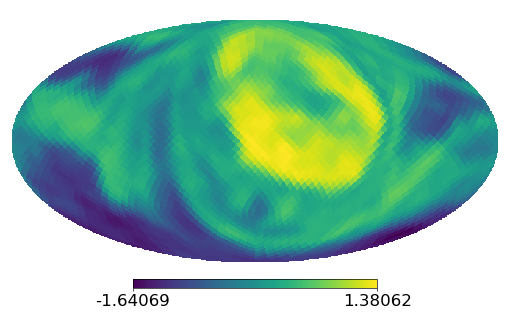}%
}\hfill
       \subfloat[ $\Delta x=0.1 \, r_g $ ]{%
  \includegraphics[width=.25\linewidth]{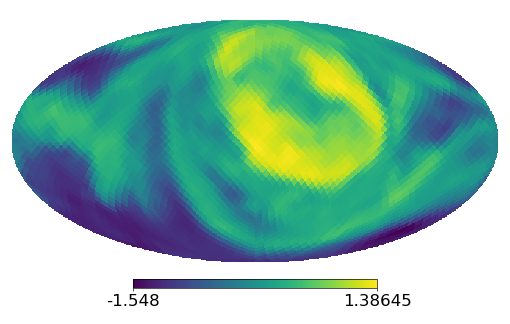}%
}\hfill
       \subfloat[ $\Delta x=0.5 \, r_g $ ]{%
  \includegraphics[width=.25\linewidth]{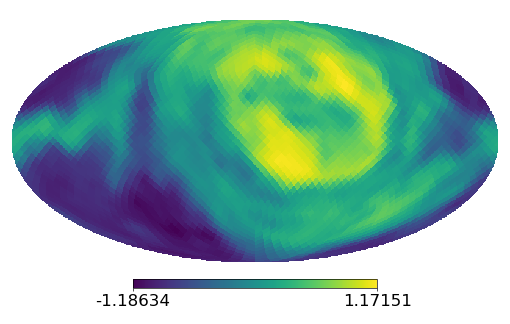}%
}\hfill
       \subfloat[ $\Delta x= 1.0 \, r_g $]{%
  \includegraphics[width=.25\linewidth]{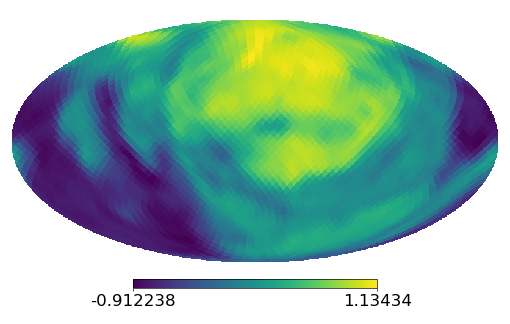}%
}\hfill
  \caption{The simulated skymap of cosmic-ray {($r_g/l_c=0.1$)} intensity for an observer at a randomly chosen reference location is shown in the left most panel. The location of the observer is then changed along a randomly chosen axis, and the simulated skymaps for varying amounts of displacement (noted in the title) along that axis are shown in the other three panels. }
\label{skymap}
\end{figure*}

\begin{figure}
  \includegraphics[width=\linewidth]{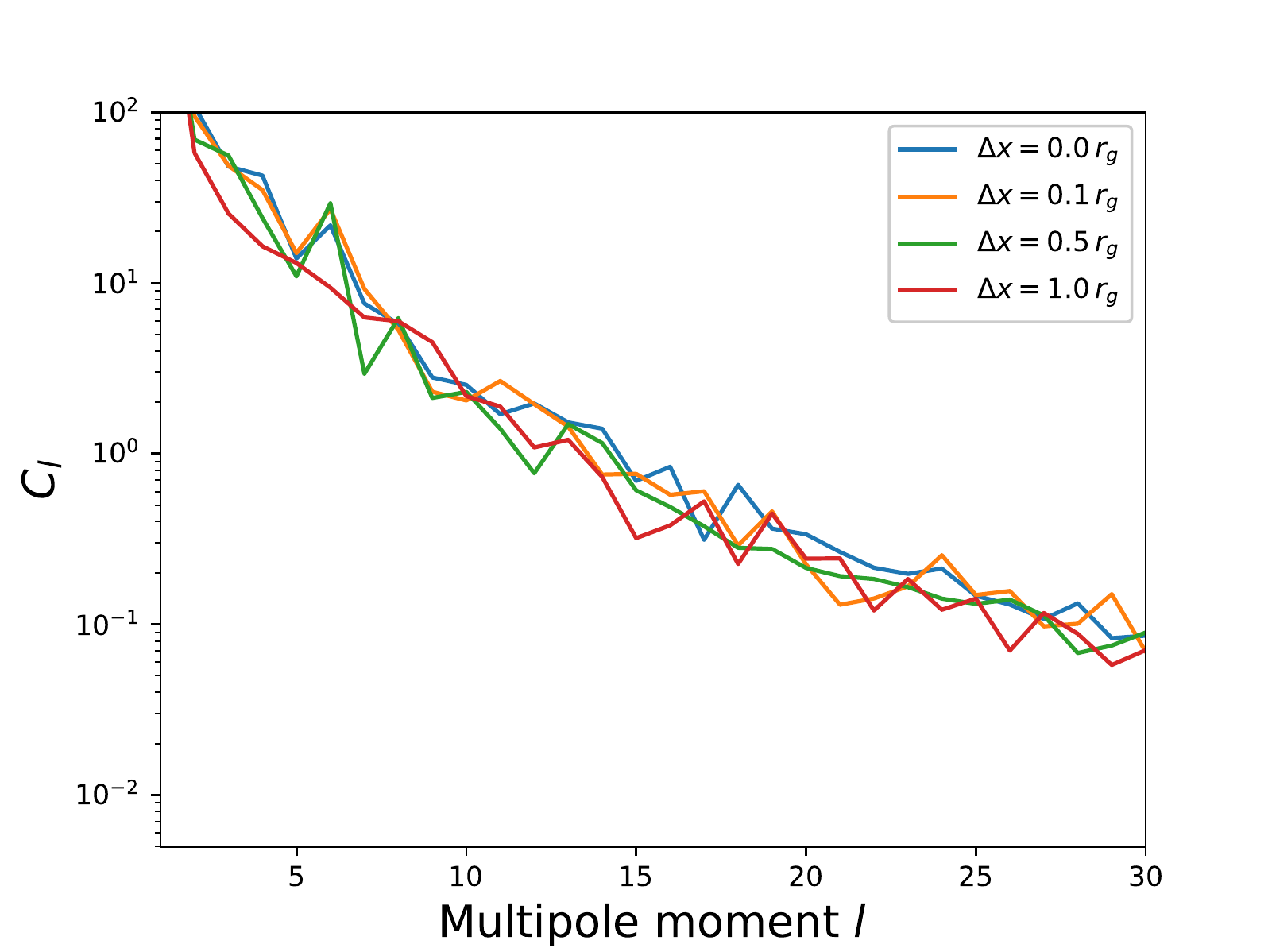}
  \caption{Power spectra of the simulated cosmic-ray skymaps at different locations of the Earth. The four different locations of the Earth shown here are the same as in the figure \ref{skymap}. }
  \label{power_spec}
\end{figure}

\begin{figure}
  \includegraphics[width=\linewidth]{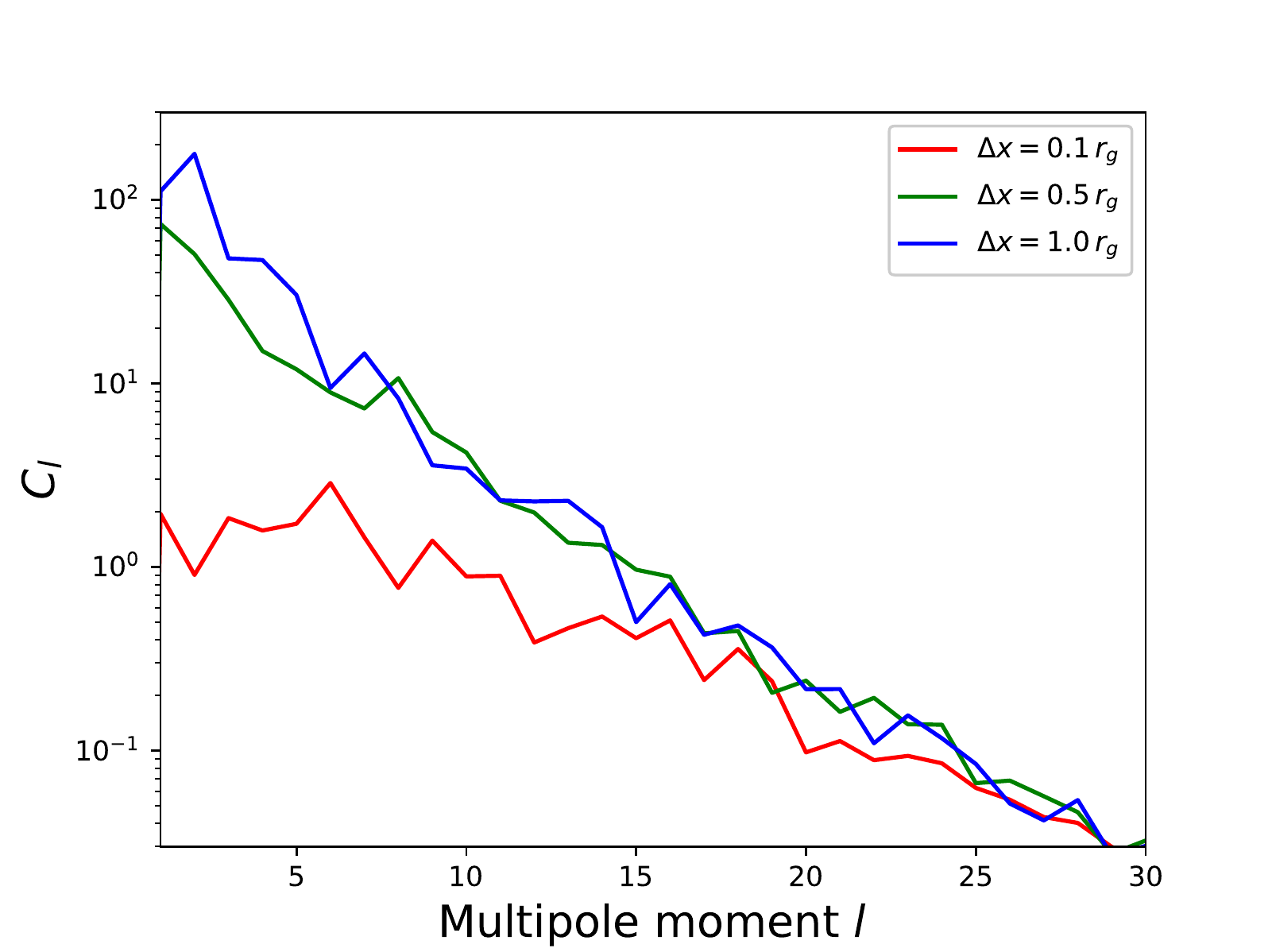}
  \caption{
\mpo{Power spectra of the variation in cosmic-ray intensity arising from relocation of the observer}, i.e. the power spectra of the difference between a skymap at the position of reference $\Delta x =0$, and three different Earth positions: $\Delta x =0.1$, $0.5$, and 1.0 $r_g$.}
  \label{diff_spec}
\end{figure}

{We vary the location of the Earth and estimate the CR flux in each angular bin, as described above. The realization of magnetic turbulence, CR density, and dipole anisotropy, as well as the size and orientation of the angular bins are kept unchanged. In figure \ref{skymap} we show the simulated cosmic-ray skymap at a randomly chosen reference point as well as at three other nearby locations. The shift in the location of the Earth, represented by $\Delta x$, \mpo{is} along a randomly chosen direction. The four example skymaps in the figure illustrate a random and appreciable variation in the observed cosmic-ray skymap for a \mpo{displacement of Earth on the order of $\sim r_g$, assuming} static magnetic turbulence. The size of angular scale affected by the change in observer's location depends on the magnitude of the change.}

{The power spectra of the CR intensity skymap at different locations of Earth are shown in figure \ref{power_spec}. As evident from the figure, the power spectrum shows only a small change due to change in the location of the Earth. The changes in the intensity skymap due to a change in the location of the Earth appears mainly as a change in the angular position of the structures in the skymap. The power spectra suggest that the amplitude of the variation in intensity at a certain angular scales remains nearly independent of our location. That is to say, the maximum variation in the intensity at certain angular scale is about the amplitude of the intensity variation at that particular angular scale in the skymap at any given location. Angular bins differ in intensity because CR in different bins have different scattering history due to interaction with different turbulent eddies with different phases and pitch angles. For a given location of the Earth the entire skymap represents an ensemble of scattering histories. In a homogeneous turbulence, the history of one angular bin is merely replaced by the other as the location of the observer changes. Therefore the power spectrum remains the same even though the intensities in individual angular bins have changed. }   

\begin{figure*}
  \centering
         \subfloat[  Reference Location  ]{%
  \includegraphics[width=.25\linewidth]{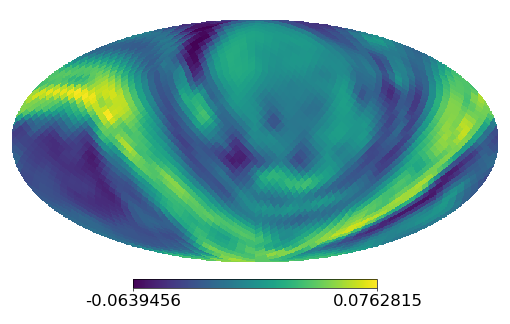}%
}\hfill
       \subfloat[ $\Delta x=1 \, r_g $ ]{%
  \includegraphics[width=.25\linewidth]{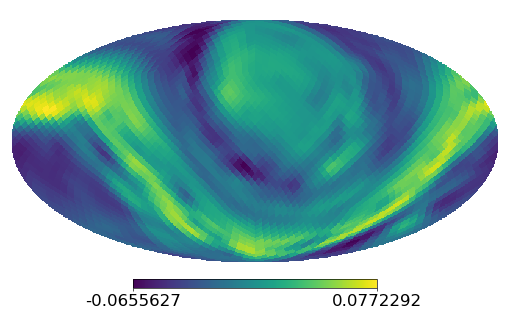}%
}\hfill
       \subfloat[ $\Delta y=1 \, r_g $ ]{%
  \includegraphics[width=.25\linewidth]{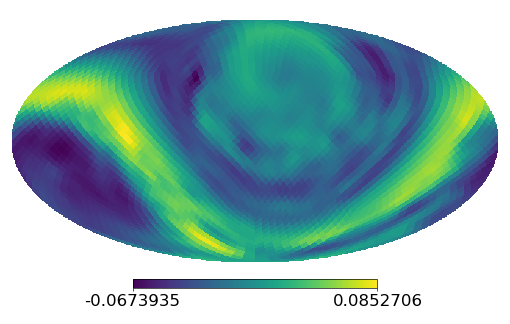}%
}\hfill
  \caption{Similar to Figure \ref{skymap}, the simulated cosmic ray intensity skymaps for $r_g=10^{-4} l_c$ are shown for three arbitrary locations of an observer. The three locations are separated by $r_g$ and the global mean magnetic field is along the x-direction. In order to emphasize the change in small angular scale structures, the dipole anisotropy is removed from all the skymaps shown in this figure.}
\label{lowEfigure}
\end{figure*}

\subsection{Temporal variation for low rigidity CRs: $r_g/l_c=10^{-4}$ }
\label{lowEsec}
The changes in the intensity skymap are due to a) alteration of the phase-space structures by the turbulent component of the magnetic field as gyrophase-bunched particles are scattered during transport from one location to the other, and b) rotation of the existing structure in the momentum space due the local mean magnetic field. A combination of both these effects gives rise to $r_g-$scale variability in the skymap. The case of $r_g=0.1 l_c$ correspond to $\sim$PeV cosmic rays and the mean free path of the CRs is only about 10 times larger than $l_c$. In order to illustrate that $r_g-$scale variation holds true for $\sim$TeV energy cosmic rays as well, we have carried out a set of three high-resolution simulation for $r_g=10^{-4} l_c$. In this case $k_{max}=2\cdot 10^5$ and about $10^7$ CR trajectories are computed to produce each skymap. The simulated skymaps for the three different locations of the observer are shown in Figure \ref{lowEfigure}. The figures show that small-scale structures in the CRs intensity skymaps change if position of an observer is changed by about $r_g$. The amplitude of intensity variation at large angular scales $(l\lesssim 5)$ is relatively smaller as compared to the $r_g/l_c=0.1$ case. 

\section{Discussion and Conclusions}

The $r_g-$scale variability in the skymap is illustrated for $r_g$ close to the correlation length of the turbulence, and therefore moderately large $\delta B/B_0$. The level of turbulence experienced by TeV particles, though smaller if the turbulent spectrum is Kolmogorov, is nevertheless unknown. It could be larger than predicted by a Kolmogorov spectrum if there are additional sources of it at smaller scales (e.g. stellar winds). In any case, a similar $r_g-$scale variability for TeV energy can be argued as follows: at a given location of the Earth different arrival directions $\hat{r}$ of cosmic rays correspond to different pitch angles and phases $\phi$ with respect to \mpo{the} local magnetic field, $B_0$, and gyroradius-scale fluctuations, $\delta B$. Under the hypothesis that the small-scale anisotropy is due to local interstellar turbulence, the variation in CR \mpo{intensity} with arrival direction \mpo{arises} because trajectories map slightly different CR \mpo{source} regions. The distinction in the history of the observed CRs is determined by their local pitch angle and phase $\phi$. A change in the location of an observer implies that the same arrival direction would correspond to different $\phi$ (possibly even \mpo{a different} pitch angle), and the exact same reason (i.e., different pitch angle and $\phi$ means different flux) would now imply a slightly different flux in the same angular bin. In other words, CRs tracing the same phase-space density have very different arrival direction for observers with gyroradius-scale separation due to the gyration of CRs in the magnetic field. 

The power spectrum of the CR intensity skymap suggests that \mpo{the} amplitude of variation in the CR intensity is a steeply increasing function of the angular scale of the anisotropy. The alteration in the relevant characteristic of CRs that determine their history are supposedly a smooth function of the arrival direction. The variation in these characteristics for the same arrival direction is expected to be proportional to the displacement of the Earth with respect to the interstellar turbulence. Therefore, a uniform motion of the Earth would imply that the variation in the CR intensity skymap at smaller angular scales would take place at a 
 shorter time scale.  In figure \ref{diff_spec} we show power spectra of the change in the intensity skymaps to illustrate the proportionality between the angular size of the anisotropy and displacement of the Earth. The change in the intensity in any differential angular bin, at a given location of the Earth, is measured with respect to the intensity skymap at a randomly chosen reference position corresponding to $\Delta x =0 $. In figure \ref{diff_spec} the red line is the power spectrum of the difference between the skymap at the position $\Delta x=0.1 \, r_g$, and the skymap at the reference position $\Delta x=0 \, r_g$.  It is evident, by comparison with the power spectrum of the CR intensity skymaps(figure \ref{power_spec}), that the amplitude of fluctuations that represent change in the intensity is much smaller for $l\lesssim 20$. That is to say, for a change of $0.1 \, r_g$ in the Earth position, the largest structures ('spots") observed on the reference map would remain nearly unaffected and only the small-scale structures (\mpo{angular scale smaller than $~10$ degrees}) would be affected (see also figure \ref{skymap}). The power spectrum of the difference in the CR intensity skymap \mpo{is} also shown for $\Delta x=0.5 \, r_g$ (in green) and $\Delta x=1 \, r_g$ (in blue). Similarly, we infer that for $\Delta x= 0.5 \, r_g$ changes in the intensity skymap are observed at \mpo{angular scales up to $~45$ degrees}. Therefore, the time scale determined by $r_g-$scale displacement of the Earth may only be considered as an upper limit on a measurable variation on the intensity skymap.




We \mpo{predict} a measurable change in the skymap of \mpo{anisotropy pattern of CRs of a few TeV in} energy in few decades if the structure are due to magnetic turbulence. Remarkably, CRs of a few TeV can also be affected by the heliosphere but any variation in the \mpo{intensity} due to heliospheric would correlate with the change in the heliosphere and would likely show an angular asymmetry in the flux variation due to the non-spherical shape of the heliospheric boundaries. {Moreover, the effect of the heliosphere should be independent of our location in the ISM and can also be better constrained with further improvements in our understanding of the global structure of the heliosphere.} Also, the structure of turbulent eddies at the gyroscale of CRs may change with time which can lead to faster change in the skymap. An Alfv\'en speed $V_A \sim10 $ km/sec would imply that the turnover time $r_g/V_A $ for eddies at gyroradius scale is comparable to the transit time of the Earth through a $r_g$-scale turbulent eddy.

The IceCube observatory has not measured any significant change in the skymap during a six years period of their operation in the Southern Hemisphere \citep{IceCube2016}. The stability of small scale anisotropy observed by IceCube could be due their higher threshold energy $(\gtrsim 10\, {\rm TeV})$ which would require a longer observational period to measure any measurable change. In any case, the limits on temporal variability of the flux skymap can potentially provide strong constraints for theoretical models concerning the origin of anisotropy. Future observations in combination with archival data \mpo{can likely shed} more light on the origin of small-scale anisotropy as the cosmic ray observatories continue to collect data.\\


We thank D. J. McComas, A. Spitkovsky, and E. J. Zirnstein for helpful discussions. We acknowledge support from the Israel-U.S. Binational Science Foundation, the Israeli Science Foundation (ISF), the ISF-University Grant Commission (India), and the Joan and Robert Arnow Chair of Theoretical Astrophysics. RK was partially supported by the Max-Planck/Princeton Center for Plasma Physics and NSF grant AST-1517638. NG acknowledges support from the Koret Foundation. M.P. acknowledges valuable discussion with the team 'The physical of the very local interstellar medium and its interaction with the heliosphere' at the International Space Science Institute in Bern, Switzerland. The analysis in this paper made use of HEALPIX package \citep{healpix}. Numerical calculations in this paper used computational resources supported by PICSciE-OIT High Performance Computing Center.


\end{document}